\documentclass[conference]{IEEEtran}
\IEEEoverridecommandlockouts
\usepackage{cite}
\usepackage{amsmath,amssymb,amsfonts}
\usepackage{algorithmic}
\usepackage{algorithm}
\usepackage{graphicx}
\usepackage{textcomp}
\usepackage{xcolor}
\def\BibTeX{{\rm B\kern-.05em{\sc i\kern-.025em b}\kern-.08em
    T\kern-.1667em\lower.7ex\hbox{E}\kern-.125emX}}
\begin{document}

\title{Data-Driven Stability Assessment of Power Electronic Converters with Multi-Resolution Dynamic Mode Decomposition\\
}

\author{\IEEEauthorblockN{1\textsuperscript{st} Rui Kong}
\IEEEauthorblockA{\textit{Department of Energy} \\
\textit{Aalborg University}\\
Aalborg, Denmark \\
ruko@energy.aau.dk}
\and
\IEEEauthorblockN{2\textsuperscript{nd} Subham Sahoo}
\IEEEauthorblockA{\textit{Department of Energy} \\
\textit{Aalborg University}\\
Aalborg, Denmark \\
sssa@energy.aau.dk}
\and
\IEEEauthorblockN{3\textsuperscript{rd} Yongjie Liu}
\IEEEauthorblockA{\textit{Department of Energy} \\
\textit{Aalborg University}\\
Aalborg, Denmark \\
yoli@energy.aau.dk}
\and
\IEEEauthorblockN{4\textsuperscript{th} Frede Blaabjerg}
\IEEEauthorblockA{\textit{Department of Energy} \\
\textit{Aalborg University}\\
Aalborg, Denmark \\
fbl@energy.aau.dk}
}

\maketitle

\begin{abstract}
Harmonic instability occurs frequently in the power electronic converter system. This paper leverages multi-resolution dynamic mode decomposition (MR-DMD) as a data-driven diagnostic tool for the system stability of power electronic converters, not requiring complex modeling and detailed control information. By combining dynamic mode decomposition (DMD) with the multi-resolution analysis used in wavelet theory, dynamic modes and eigenvalues can be identified at different decomposition levels and time scales with the MR-DMD algorithm, thereby allowing for handling datasets with transient time behaviors, which is not achievable using conventional DMD. Further, the selection criteria for important parameters in MR-DMD are clearly defined through derivation, elucidating the reason for enabling it to extract eigenvalues within different frequency ranges. Finally, the analysis results are verified using the dataset collected from the experimental platform of a low-frequency oscillation scenario in electrified railways featuring a single-phase converter.
\end{abstract}


\section{Introduction}
With the advent of modern power electronic-based power systems involving widespread applications of grid-tied power electronic converters, there are frequent occurrences of various small-signal stability issues across a wide frequency range \cite{r1}. There are two main analytical methods, i.e., eigenvalue analysis based on the state-space model \cite{r2} and the impedance-based analysis based on the transfer functions \cite{r3}. However, due to the complexity and limited controller information of modeling actual power electronic converter systems, data-driven mode identification techniques are more practical. 

Regarding mode identification methods, Prony algorithm\cite{r4}, Matrix Pencil algorithm\cite{r5}, and eigensystem realization algorithm (ERA) \cite{r6} are reported and compared in \cite{r7}, which can extract system eigenvalues to assess system stability intuitively. As a powerful data-driven tool originating from the field of fluid dynamics, dynamic mode decomposition (DMD) \cite{r8} can extract dynamic modes and eigenvalues for stability assessment by performing eigen-decomposition of the best fit dynamic matrix dimensionally reduced via singular value decomposition (SVD) \cite{r9}. With unique advantages in capturing spatio-temporal characteristics of data snapshots, DMD has been used for oscillation analysis in power systems \cite{r10}, and it was demonstrated that DMD has clear physical significance and superior accuracy compared to the aforementioned three identification methods \cite{r10}. However, DMD performs global processing on all data within the sampling time window, which results in poor robustness to correctly identify transient time behaviors or outliers within the dataset such as stability state changing and missing data \cite{r11}.

To address the aforementioned problems, an oscillation identification method based on multi-resolution dynamic mode decomposition (MR-DMD) \cite{r11} is presented in this paper. Dominant dynamic modes can be obtained in different frequency ranges and varying timescales for stability assessment. Then, selection criteria for the key parameters of MR-DMD are developed through rigorous mathematical derivations. The method performance and its advantages are demonstrated with an experimental dataset sampled from a single-phase converter platform when low-frequency oscillation occurs, but the proposed methodology is generic for wide frequency ranges. 

\begin{figure}[htbp]
\centering
\includegraphics[width=0.89\columnwidth]{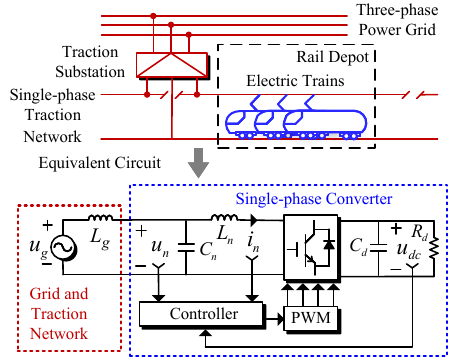}
\caption{System diagram of a single-phase converter in electrified railways.}
\label{System_structure}
\end{figure}

\section{System Description amd Problem Formulation}
\subsection{System Description}

\begin{figure}[htbp]
\centering
\includegraphics[width=0.95\columnwidth]{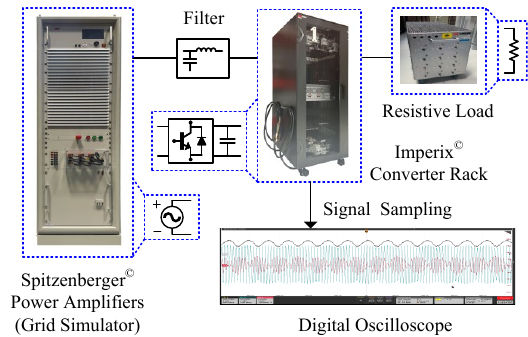}
\caption{Experimental setup of a down-scaled single-phase converter in an electrified railway system.}
\label{Experiment_Setup}
\end{figure}

Considering a single-phase converter of electric trains as shown in Fig. \ref{System_structure}, low-frequency oscillations (LFO) might occur in electrified railways when multiple trains are energized in the same rail depot under low power conditions \cite{r12}. When LFOs occur, there are periodic oscillations of AC-side voltage and current, and DC-side voltage with a frequency of 1 to 10 Hz \cite{r12}. The transient direct current control strategy \cite{r13} is used in the controller, but the measured signals are independent of the controller due to the assumption of black-box conditions with an unknown controller structure. To assess system stability with data-driven methods, the dataset of system states is collected from an experimental platform of the downscaled single-phase converter as shown in Fig. \ref{Experiment_Setup}, and Table \ref{Experiment_parameters} lists the main system parameters. 

\begin{table}[htbp] 
\centering
\caption{Main Parameters of Experimental System}
\renewcommand{\arraystretch}{1.6}
\begin{tabular}{ccc} 
\hline\hline
\textbf{Parameter} & \textbf{Description}                                                                              & \textbf{Value}  \\ 
\hline
${u_g}$             &  Grid phase voltage (RMS)                                                               & 110  V        \\ \hline
${L_g}$             & Grid-side inductance                                                                      & 8  mH      \\ \hline
${L_n}$             & Converter-side inductance                                                                  & 4  mH        \\ \hline
${C_n}$             & Filter capacitance                                                                 & 10   uF        \\ \hline
${u_{dc}}$             & DC link voltage of the converter                                                   & 170  V        \\ \hline
${R_d}$             & DC link Load resistance                                                                     & 460 $\Omega$       \\ \hline
${C_d}$              & DC link support capacitance                                                                 & 800  uF          \\ \hline
${f_s}$               & Switching frequency                                                                         & 10  kHz        \\ \hline
\hline
\end{tabular}
\label{Experiment_parameters}
\end{table}

\subsection{DMD Algorithm}
\begin{figure*}[htbp]
\centering
\includegraphics[width=2 \columnwidth]{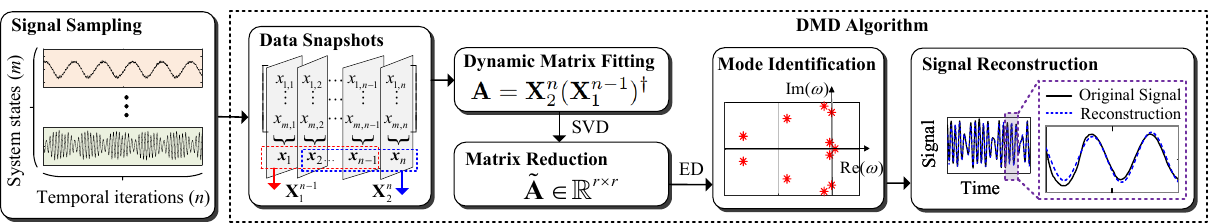}
\caption{Schematic of dynamic mode decomposition (DMD) (SVD: singular value decomposition, ED: eigen-decomposition).}
\label{DMD_1}
\vspace{-3pt}
\end{figure*}

Based on the spatio-temporal snapshots of discrete measurement data, DMD can find a best-fit solution of the dynamic matrix ${\bf{A}}$ for oscillation mode identification as shown in Fig. \ref{DMD_1}, where dynamic matrix  ${\bf{A}} \in {\mathbb{R}^{m \times m}}$ is the mapping relationship between a pair of data matrices with time shift, governing the inherent dynamic evolution of a discrete linear system \cite{r8}. In Fig. \ref{DMD_1}, ${x_{i,j (i = 1,2, \ldots, m;\ j = 1,2, \ldots, n) }}$ denotes an element of the measurable discrete system states, where $i$ is the index of system states, $m$ is the total number of system states, $j$ is the index of temporal iterations, and $n$ is the total number of temporal iterations with regular sampling time interval $\Delta t$. 

The standard DMD procedure \cite{r10} is summarized in Algorithm \ref{alg_DMD}, which starts with generating a pair of data matrices, i.e., ${\bf{X}}_1^{n - 1} \in {\mathbb{R}^{m \times (n - 1)}}$ and ${\bf{X}}_2^n \in {\mathbb{R}^{m \times (n - 1)}}$ with a time shift, and ends up with the signal reconstruction ${\bf{X}} ^{rec} \in {\mathbb{R}^{m \times n }}$. It is worth mentioning that, since high order data matrices are required in the DMD algorithm but there are limited measurable channels in power converter systems, the data-stacking technique \cite{r14} needs to be used before executing the algorithm to increase the row number ($m$) of the data matrix and ensure all significant modes are contained.

\begin{algorithm}[H]
\caption{ Dynamic Mode Decomposition (DMD)}\label{alg:alg1}
\begin{algorithmic}
\STATE 
\STATE \textbf{Input:} ${\mathop{\bf{X}} \nolimits} _1^n{\rm{ = [}}\begin{array}{*{20}{c}}{{{\mathop{\bf{x}} \nolimits} _1}}&{{{\mathop{\bf{x}} \nolimits} _2}}& \cdots & {{{\mathop{\bf{x}}\nolimits} _{n - 1}}}&{{{\mathop{\bf{x}}\nolimits} _n}}\end{array}{\rm{]}} \in {\mathbb{R}^{m \times n}}$
\vspace{0.6ex}
\STATE \textbf{Output:} ${\bf{\Phi }}\in {\mathbb{R}^{m \times r}}$, ${\bf{\Lambda }}\in {\mathbb{R}^{r \times r}}$, ${\bf{b}}\in {\mathbb{R}^{r \times 1}}$, ${{\bf{x}} _j^{rec}}\in {\mathbb{R}^{m \times 1}}$ 

\STATE \textbf{Step:}
\STATE 1: \hspace{0.02cm} ${\mathop{\bf{X}}\nolimits} _1^{n-1}{\rm{ = [}}\begin{array}{*{20}{c}} {{{\mathop{\bf{x}}\nolimits} _1}}&{{{\mathop{\bf{x}}\nolimits} _2}}& \cdots &{{{\mathop{\bf{x}}\nolimits} _{n-1}}} \end{array}{\rm{]}} \in {\mathbb{R}^{m \times {(n-1)}}}$
\STATE\hspace{0.4cm} ${\mathop{\bf{X}}\nolimits} _2^n{\rm{ = [}}\begin{array}{*{20}{c}} {{{\mathop{\bf{x}}\nolimits} _2}}&{{{\mathop{\bf{x}}\nolimits} _3}}& \cdots &{{{\mathop{\bf{x}}\nolimits} _n}} \end{array}{\rm{]}} \in {\mathbb{R}^{m \times {(n-1)}}}$
\STATE\hspace{0.4cm} ${\rm{ }}{\bf{A}} = {\bf{X}}_2^n{({\bf{X}}_1^{n - 1})^\dag \in {\mathbb{R}^{m \times m}}}$
\STATE 2: \hspace{0.02cm} $ {\rm{[}}{\bf{U}}, {\bf{\Sigma}}, {\bf{V}}{\rm{]}} = {\rm{SVD}}({\mathop{\bf{X}}\nolimits} _1^{n - 1})  \leftarrow  {\bf{X}}_1^{n - 1} \approx {\bf{U\Sigma }}{{\bf{V}}^T}$
\STATE\hspace{0.6cm} ${\bf{U}}\in {\mathbb{R}^{m \times r}}$, ${\bf{\Sigma }}\in {\mathbb{R}^{r \times r}}$, ${\bf{V}}\in {\mathbb{R}^{(n-1) \times r}}$
\STATE 3: \hspace{0.02cm} ${\bf{\tilde A}} = {{\bf{U}}^T}{\bf{AU}} = {{\bf{U}}^T}{\bf{X}}_2^n{\bf{V}}{{\bf{\Sigma }}^{ - 1}}$, ${\bf{\tilde A}} \in {\mathbb{R}^{r \times r}}$
\vspace{0.5ex}
\STATE 4: \hspace{0.02cm} ${\rm{[}}{\bf{W}}, {\bf{\Lambda }}{\rm{]}} = {\rm{ED}}({\bf{\tilde A}}) \leftarrow {\bf{\tilde A}}{{\bf{W}}} = {\bf{W\Lambda }}$    
\STATE 5: \hspace{0.02cm} ${\bf{\Phi }} = {\bf{U}}{\bf{W}} \leftarrow {\bf{A}}{{\bf{\Phi}}} = {\bf{\Phi\Lambda }} $   
\STATE 6: \hspace{0.02cm} ${\bf{b}} = {{\bf{\Phi }}^{\dag}}{{\bf{x}}_1}$
\vspace{-0.6ex}
\STATE 7: \hspace{0.02cm} ${{\bf{x}}_{j}^{rec}} = {\bf{\Phi }}{{\bf{\Lambda }}^{j-1}}{{\bf{\Phi }}^{\dag}}{{\bf{x}}_1} = {\bf{\Phi }}{{\bf{\Lambda }}^{j-1}}{\bf{b}} = \sum\limits_{k = 1}^r {{{\bf{\Phi }}_k}} {({\lambda _k})^{j-1}}{b_k} $  
\STATE\hspace{0.4cm} ${\mathop{\bf{X}} \nolimits} ^{rec}{\rm{ = [}}\begin{array}{*{20}{c}}{{{\mathop{\bf{x}} \nolimits} _1}} & \cdots & {{{\mathop{\bf{x}}\nolimits} _j^{rec}}}& \cdots &{{{\mathop{\bf{x}}\nolimits} _n^{rec}}}\end{array}{\rm{]}} \in {\mathbb{R}^{m \times n}}$
\vspace{-0.6ex}
\end{algorithmic}
\label{alg_DMD}
\end{algorithm}

In the DMD algorithm, the dynamic matrix ${\bf{A}}$ is obtained in step 1, where $\dag $ denotes the Moore-Penrose pseudo-inverse.  Then, singular value decomposition (SVD) with the truncation rank $r$ ${\rm{ }}(r \ll m)$ is implemented to calculate the pseudo-inverse of ${\bf{X}}_1^{n - 1}$, and a reduced-order matrix ${\bf{\tilde A}}$ with rank $r$ can be obtained by projecting the initial full dynamic matrix ${\bf{A}}$ into an ${\bf{U}}$ basis subspace. Then, the eigen-decomposition (ED) of ${\bf{\tilde A}}$ is calculated to extract diagonal matrix ${\bf{\Lambda }} = diag\left[ {\begin{array}{*{20}{c}} {{\lambda _1}}&{{\lambda _2}}& \cdots &{{\lambda _r}} \end{array}} \right]$ consisting of eigenvalues, while matrix of DMD exact modes ${\bf{\Phi }}$ is obtained based on eigenvectors ${\bf{W}}$. Finally, the measured signal (data matrix) can be reconstructed, where ${\bf{b}}$ is defined as the mode amplitude, ${{\bf{\Phi }}_k}$ is the $k^\text{th}$ column of matrix ${\bf{\Phi }}$, ${b_k}$ is the $k^\text{th}$ mode amplitude with $k=1,2,...,r$.

\begin{figure}[htbp]
\centering
\includegraphics[width=0.98\columnwidth]{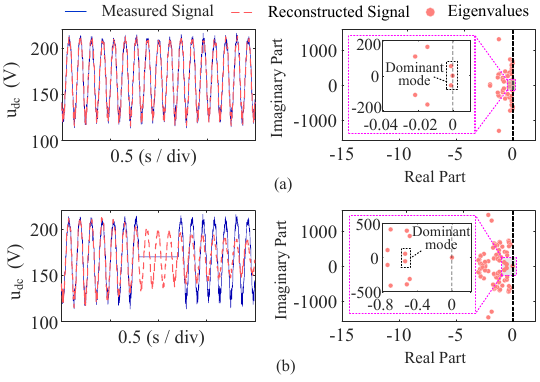}
\caption{Measured signal reconstruction and eigenvalues identification by DMD using (a) normal dataset and (b) dataset with missing data — DMD fails to accurately assess stability based on the latter.}
\label{DMD_for_Comparison_2}
\end{figure}

However, DMD just deals with all data within the sampling window at once to provide a full-rank approximation of the dynamics observed. As an SVD-based algorithm, a single DMD application can only yield one pair of eigenvalues for the same mode, resulting in poor robustness to correctly capture transient time behaviors within the dataset, which can be observed clearly by the comparison shown in Fig. \ref{DMD_for_Comparison_2}.

As shown in Fig. \ref{DMD_for_Comparison_2}(a), DMD is capable of reconstructing measured signals and identifying dominant eigenvalues with critical damping, where DC-side voltage ${u_{dc}}$ is employed as the measured signal to obtain the dataset when the LFO at 8.6 Hz occurs in the experimental platform as shown in Fig. \ref{Experiment_Setup}. However, it can be seen from Fig. \ref{DMD_for_Comparison_2}(b) that, if there is transient time behavior in the signal such as missing data, DMD will fail to perform well. To be specific, there is a large error in the damping of dominant eigenvalues, showing incorrect stability assessment results. Moreover, the measured signal cannot be reconstructed accurately.

\section{Mode Identification with MR-DMD}
\subsection{MR-DMD Algorithm}
\begin{figure}[htbp]
\centering
\includegraphics[width=0.98\columnwidth]{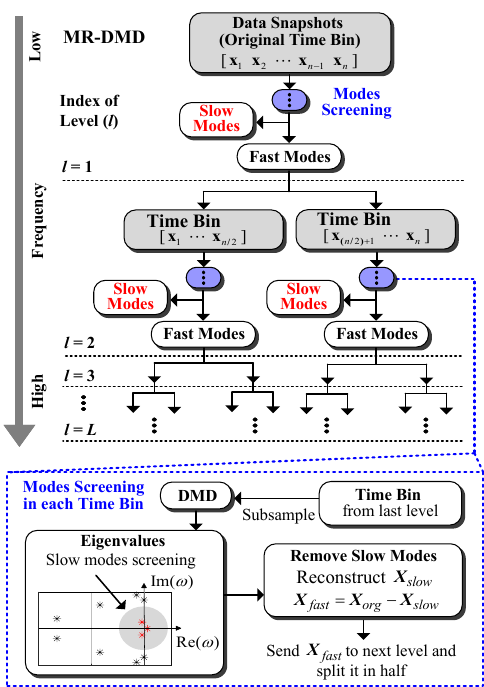}
\caption{Schematic of multi-resolution dynamic mode decomposition (MR-DMD) — DMD is performed recursively in varying time scales and slow modes are screened in each level.}
\label{MR_DMD}
\end{figure}

By integrating the conventional DMD algorithm with multi-resolution analysis from wavelet theory \cite{r15}, multi-resolution dynamic mode decomposition (MR-DMD) applies DMD recursively in different spatio-temporal subsamples \cite{r11}. In the MR-DMD algorithm, oscillation modes with relatively low frequency and slow growth/decay rate are defined as slow modes, so their corresponding eigenvalues are close to the origin of the complex plane. As shown in Fig. \ref{MR_DMD}, slow modes can be screened after performing the DMD algorithm on the available dataset (i.e., original time bin), and then the DMD approximation ${\bf{X}}_{slow}$ is found from only the slow modes. After that, the slow-mode approximation ${\bf{X}}_{slow}$ is removed from the available dataset, and the remaining part ${\bf{X}}_{fast}$ representing fast modes is split in half to obtain two equal time bins. The above steps are repeated recursively for each time bin in different decomposition levels until a desired truncation level is achieved. It is worth noting that, since only the slow modes at each decomposition level are captured, it is important to subsample with a fixed small number in each time bin, which can not only improve the computational efficiency but also ensure the maximum frequency of capturable mode increase with increasing decomposition level. 

\subsection{Parameter Derivation }
According to the steps of MR-DMD as shown in Fig. \ref{MR_DMD}, three important parameters affecting algorithm performance need to be clearly defined, i.e., the fixed subsample number of each time bin ($\mu $), the termination level ($L$), and the screening threshold of slow modes ($\rho$).

The recursion process of MR-DMD shows that, if the index of decomposition levels is defined as $l = 1,2,..., L$, the number of time bins in different levels can be expressed as $B = {2^{(l - 1)}}$, so the size of each time bin in different levels will be $S = n/B$, where $n$ has been defined as the total number of initial sampling points. Since the state signal is sampled with a regular time interval $\Delta t$, there is $\Delta t = N/n$, where $N$ is the sampling window duration of the original dataset. Therefore, the time window duration of each time bin ($D$) is derived as:

\begin{equation}
\label{time window duration of each time bin}
{D} = \frac{N}{{{B}}} = \frac{S N}{{{n}}} = {S}\Delta t
\end{equation}

Subsampling with a fixed number $\mu$ is indispensable before implementing DMD in each time bin, so the subsample frequency is obtained by calculating the reciprocal of the subsampling interval as:
\begin{equation}
\label{subsample frequency}
{f_{sp}} = \frac{{\mu}}{D} = \frac{\mu }{{S\Delta t}}
\end{equation}

Based on the Nyquist sampling theorem, the maximum frequency of capturable modes ${f_m}$ is obtained as:
\begin{equation}
\label{maximum frequency of capturable modes in the lth level}
f_m = \frac{{f_{sp}}}{2} = \frac{\mu }{{2{S}\Delta t}} = {2^{(l - 2)}}\frac{\mu }{N}
\end{equation}

Since $N$ has been determined after obtaining the original dataset, it can be seen from \eqref{maximum frequency of capturable modes in the lth level} that, ${f_m}$ depends on $l$ and $\mu$, which explains why the maximum frequency of capturable mode increase with increasing decomposition level after performing subsampling. Moreover, $\mu $ can be determined based on required ${f_m}$ for each decomposition level. 

The size of time bins in the highest level should be larger than the subsample number $\mu $ of each time bin, otherwise subsampling will fail to perform. Therefore, the selection criterion for the termination level $L $ is expressed as: 
\begin{equation}
\label{termination level}
\frac{n}{{{2^{(L - 1)}}}} > \mu 
\end{equation}

Hence, $L$ is affected by $\mu $. To be specific, $L$ has to be decreased if $\mu $ increases to some extent. 

Regarding screening of slow modes, When a discrete eigenvalue $\lambda_k $ identified by MR-DMD enable $\left| {\ln (\lambda_k )} \right| < \rho $ to hold, the corresponding mode can be regarded as slow mode, but screening threshold of slow modes $\rho $ is still not defined clearly. To this end, we assume the relationship between ${f_m}$ and the maximum frequency of slow modes ${f_{slow}}$ in each decomposition level is ${f_{slow}} \le {f_m}/g{\rm{ }}$, which can be further written as: 
\begin{equation}
\label{Im and fm}
\frac{{\left| {{\mathop{\rm Im}\nolimits} [{\omega _{slow}}]} \right|}}{{2\pi }} < \frac{{{f_m}}}{g}
\end{equation}
Where, $g$ is an arbitrary constant rational number larger than 1, and $\omega $ represents continuous eigenvalue in the complex plane. 

The relationship between the discrete eigenvalue $\lambda $ and continuous eigenvalue $\omega $ is $\omega  = \ln (\lambda )/\Delta t$ \cite{r10}, which can be rewritten after subsampling as:
\begin{equation}
\label{further relationship between the discrete and continuous eigenvalue}
\omega  = \frac{{\ln ({\lambda ^{\mu /s}})}}{{\Delta t}} = \frac{{\mu \ln (\lambda )}}{{s\Delta t}} = 2{f_m}\ln (\lambda )
\end{equation}

The key eigenvalues corresponding to slow modes are usually located near the imaginary axis, indicating their real parts are much smaller than imaginary parts, so $\left| {{\mathop{\rm Im}\nolimits} [{\omega _{slow}}]} \right| \approx \left| {{\omega _{slow}}} \right|$ is assumed to be valid. As a result, $\left| {\ln ({\lambda _{slow}})} \right| < \pi /g$ is obtained by combining \eqref{Im and fm} and \eqref{further relationship between the discrete and continuous eigenvalue}, thereby $\pi /g$ can be regarded as the screening threshold of slow modes $\rho $.

\section{Experimental Results}
\subsection{Data Sampling and Processing}
Based on the experimental platform as shown in Fig. 2, the dataset is collected in a sampling window of 2 seconds with a frequency of 2500 Hz, respectively using DC-side voltage ${u_{dc}}$ and AC-side current ${i_{n}}$ as the measured signal in two cases. Hence, the total number of sampling points is 5000. Data stacking technique is employed to preprocess the data and stacking number is set as 1000, ending up with a data matrix of 1000 rows by 4000 columns $(m = 1000,{\rm{ }}n = 4000)$, so the sampling window duration $N$ is considered as 1.6 seconds. To verify robustness, the dataset was set up with missing data.

\subsection{Parameter Setting and Results}
\subsubsection {DC-side voltage ${u_{dc}}$ as measured signal}
The frequency of oscillation component in ${u_{dc}}$ denoted as $f_{LFO}$ is usually lower than 10 Hz. According to \eqref{maximum frequency of capturable modes in the lth level} and related discussion, it can be found that a smaller $\mu $ leads to more decomposition levels in the low-frequency range, allowing for finer identification of LFO. Therefore, subsample number $\mu $ of each time bin is set as 16, indicating that the maximum frequency of capturable modes ${f_m}$ in each level is expected to be $5 \times {2^{(l - 1)}}$ Hz $(l = 1, 2, ..., L)$, where the termination level $L$ can be calculated as 8 based on \eqref{termination level}. The screening threshold of slow modes $\rho $ is set as $\pi /4$, meaning that modes with frequencies less than ${f_m}/4$ in each level are regarded as slow modes. Hence, the maximum frequency of slow modes ${f_{slow}}$ in each level should be $5 \times {2^{(l - 3)}}$ Hz.

\begin{figure*}[htbp]
\centering
\includegraphics[width=1.98\columnwidth]{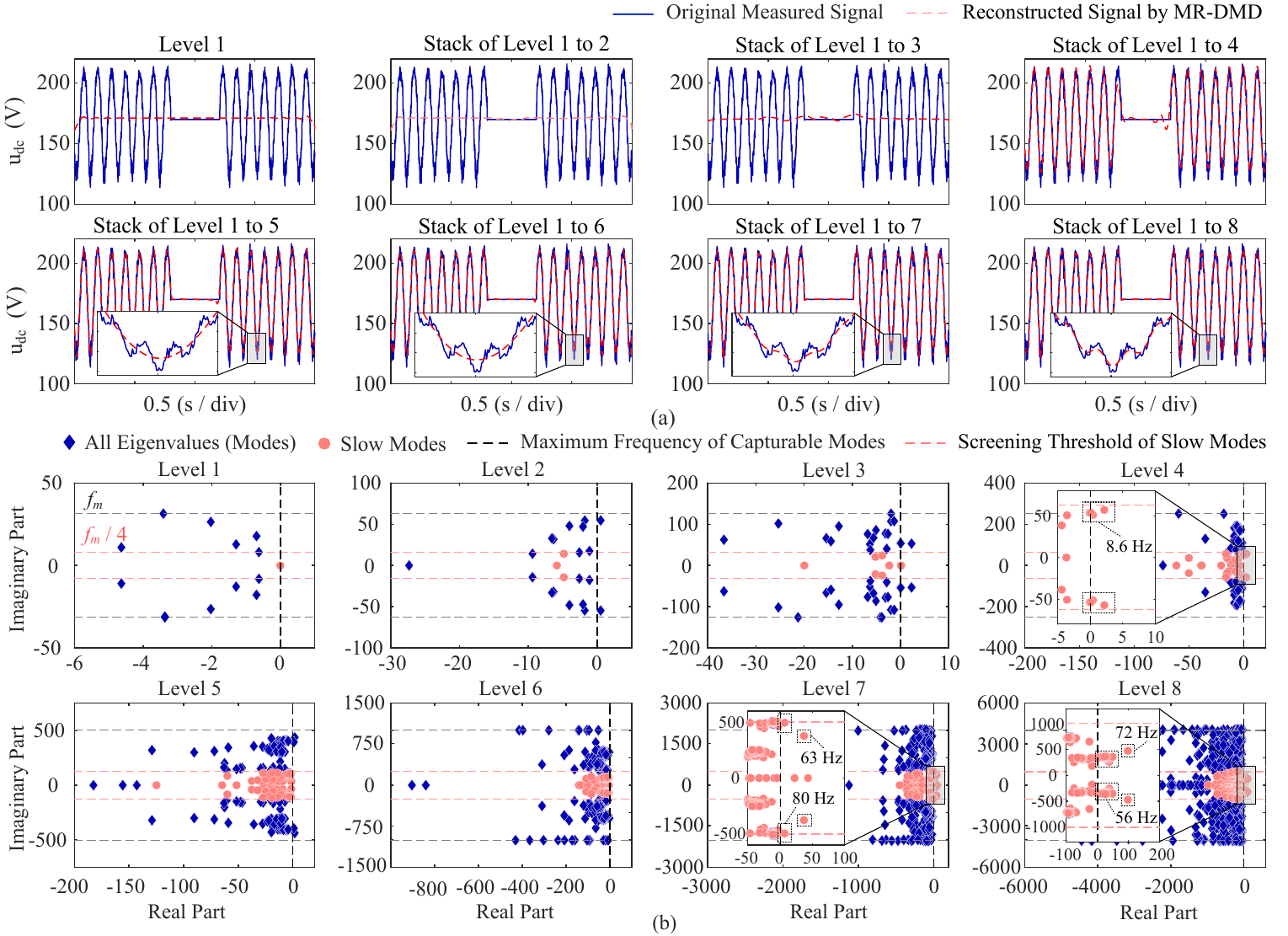}
\caption{Identification results by MR-DMD using ${u_{dc}}$ as measured signal (a) signals reconstructions, (b) eigenvalues analysis — Measured signal is reconstructed accurately and eigenvalues with different frequency ranges are identified in each decomposition level.}
\label{Eigenvalue_and_reconstruction_DC}
\vspace{-10pt}
\end{figure*}

As shown in Fig. \ref{Eigenvalue_and_reconstruction_DC}(a), by superimposing all 8 decomposition levels, the final reconstructed signal aligns closely with the original measured signal showing oscillations at approximately 8.6 Hz. It can also be found that Level 1 determines the basis of the signal (DC component), while Level 4 serves a significant role in identifying the low-frequency oscillation, and higher levels provide further fine fitting of the higher harmonics. All identified eigenvalues and eigenvalues of slow modes in each decomposition level are shown in Fig. \ref{Eigenvalue_and_reconstruction_DC}(b), where ${f_m}$ increases from 5 Hz in Level 1 to 640 Hz in Level 8. Since the screening threshold $\rho $ is set as $\pi /4$, slow modes marked by red points are screened in the frequency range from less than 1.25 Hz in Level 1 to less than 160 Hz in Level 8, which is consistent with the ${f_{slow}}$ designed through parameter setting.

\begin{figure}[htbp]
\centering
\includegraphics[width=0.98\columnwidth]{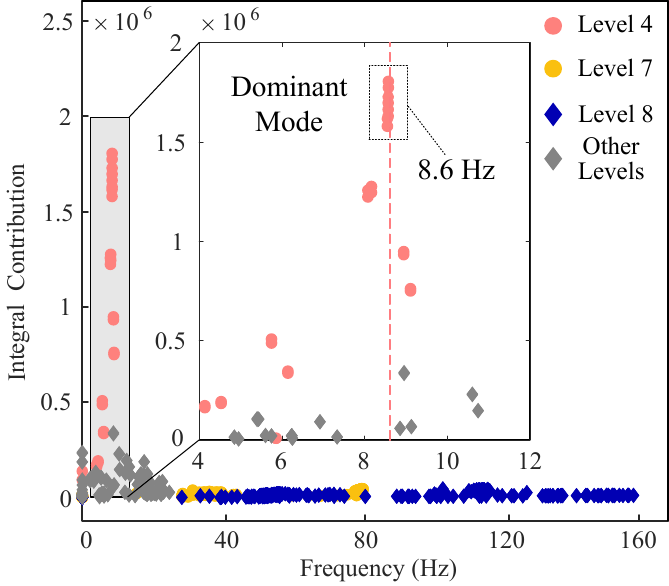}
\caption{Integral contribution of each eigenvalue — The mode with a frequency of 8.6 Hz in Level 4 is the dominant mode.}
\label{IC_DC}
\vspace{-15pt}
\end{figure}

In addition, eigenvalues (modes) with critical or negative damping are shown in Level 4, 7, and 8, but the integral contribution representing the importance of each eigenvalue \cite{r16} shown in Fig. \ref{IC_DC} indicates that the mode with a frequency of 8.6 Hz in Level 4 is the dominant mode, matching the results in Fig. \ref{Eigenvalue_and_reconstruction_DC}(a). More importantly, the above results prove that MR-DMD is more robust and can correctly handle the dataset with missing data due to its ability of multi-resolution analysis, which cannot be achieved by conventional DMD.

\subsubsection {AC-side current ${i_{n}}$ as measured signal} 
When LFO occurs, the frequency transformation of ac-dc converters translates the oscillation component at the dc-side into two components of the frequencies $ f_{o} \pm f_{LFO}$ at the ac-side, where $f_{o}$ is the grid fundamental frequency (50 Hz). Thus, $\mu $ can be set as 50 in this case, and the corresponding ${f_m}$ in each level is calculated as $15.625 \times {2^{(l - 1)}}$ Hz $(l = 1, 2, ..., L)$, where $L$ will be 6. Finally, ${f_{slow}}$ in each level should be $15.625 \times {2^{(l - 3)}}$ Hz, when $\rho $ is also set as $\pi /4$.

\begin{figure}[htbp]
\centering
\includegraphics[width=0.98\columnwidth]{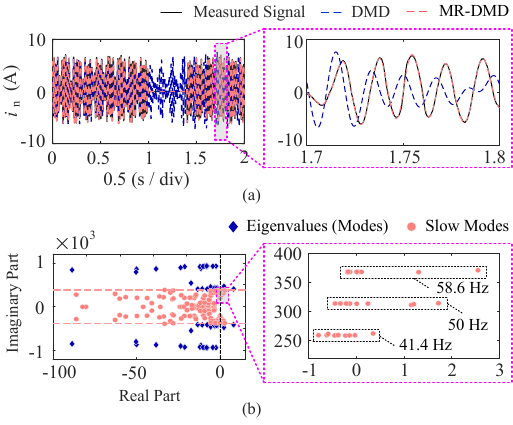}
\vspace{-6pt}
\caption{Identification resultsusing ${i_{n}}$ as measured signal (a) Comparison of reconstructed signals, (b) Identified eigenvalues in Level 5 — MR-DMD performs better than DMD when there is missing date in the dataset.}
\label{Eigenvalue_and_reconstruction_AC}
\end{figure}

As shown in Fig. \ref{Eigenvalue_and_reconstruction_AC}(a), conventional DMD fails to reconstruct measured siganl, but the final reconstructed signal of MR-DMD by superimposing all 6 decomposition levels aligns closely with the measured signal, and the envelope of the waveform shows an oscillation at 8.6 Hz. The reconstructed signals in each level will not be shown for brevity, but it is worth mentioning that the reconstruction of AC waveforms near the fundamental frequency is realized in Level 5, which is consistent with the calculation result of ${f_{slow}}$. 

\begin{figure}[htbp]
\centering
\includegraphics[width=0.98\columnwidth]{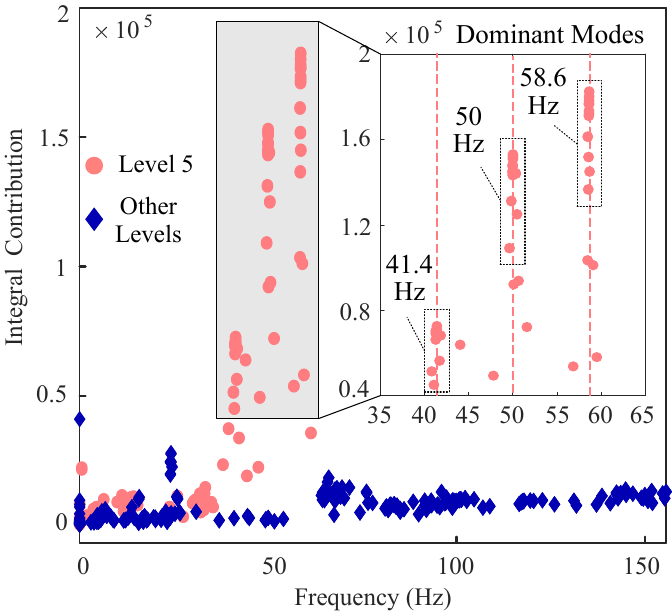}
\vspace{-6pt}
\caption{Integral contribution of each eigenvalue — The modes with frequencies of 41.4 Hz and 58.6 Hz in Level 5 are dominant modes.}
\label{IC_AC}
\end{figure}

Moreover, identified eigenvalues in Level 5 by MR-DMD are shown in Fig. \ref{Eigenvalue_and_reconstruction_AC}(b). In addition to the fundamental frequency component at 50 Hz, there are two groups of dominant modes with critical or negative damping at 41.4 Hz and 58.6 Hz respectively, matching the results of integral contribution as shown in Fig. \ref{IC_AC}. The above analysis can also accurately identify the occurrence of LFO at 8.6 Hz in the system.

\section{Conclusion}
In this paper, a diagnostic and identification tool for system stability of power electronic converters is proposed based on multi-resolution dynamic mode decomposition (MR-DMD), which hierarchically decomposes datasets into multiple time bins and applies conventional DMD recursively, where slow modes in different decomposition levels and time scales are screened. Three crucial parameters of MR-DMD are derived to reveal the frequency range of eigenvalues captured in each decomposition level and the screening threshold for slow modes. The identification results indicate that, compared with the conventional DMD algorithm, MR-DMD can accurately identify dominant oscillation modes and reconstruct measured signals using a dataset with transient time behavior. To extend the scope of this article in the future, the influence mechanism and design framework of algorithm parameters can be further analyzed to achieve optimal identification.


\bibliographystyle{IEEEtran}
\bibliography{Mybib}

\end{document}